\documentclass[10pt,letterpaper]{article}
\usepackage{amsmath,amssymb}
\usepackage{opex3}
\usepackage{cite}

\begin{document}

%%%%%%%%%%%%%%%%%% title page information %%%%%%%%%%%%%%%%%%

\title{Generating superposition of up-to three photons for continuous variable quantum information processing}

\author{Mitsuyoshi Yukawa,$^{1}$ Kazunori Miyata,$^{1}$ Takahiro Mizuta,$^{1}$ Hidehiro~Yonezawa,$^{1}$ Petr Marek,$^{2}$\\Radim Filip,$^{2}$ and Akira Furusawa$^{1,*}$}

\address{$^{1}$Department of Applied Physics, School of Engineering, The University of Tokyo,\\ 7-3-1 Hongo, Bunkyo-ku, Tokyo 113-8656, Japan\\$^{2}$Department of Optics, Palack\'y University, 17. listopadu 1192/12,\\ 77146 Olomouc, Czech Republic}

\email{$^{*}$akiraf@ap.t.u-tokyo.ac.jp}

\begin{abstract}
We develop an experimental scheme based on a continuous-wave (cw) laser for generating arbitrary superpositions of photon number states. In this experiment, we successfully generate superposition states of zero to three photons, namely advanced versions of superpositions of two and three coherent states.
They are fully compatible with developed quantum teleportation and measurement-based quantum operations with cw lasers. Due to achieved high detection efficiency, we observe, without any loss correction, multiple areas of negativity of Wigner function, which confirm strongly nonclassical nature of the generated states.
\end{abstract}


%%%%%%%%%%%%%%%%%%%%%%% References %%%%%%%%%%%%%%%%%%%%%%%%%

%%%%%%%%%%%%%%%%%%%%%%%%%%  body  %%%%%%%%%%%%%%%%%%%%%%%%%%
\section{Introduction}
Quantum information processing (QIP) has dramatically changed the way we view information. By tying it firmly with physical systems, the bits and pieces of information ceased to be theoretical constructs and became bound by their carriers' physical properties. In some areas this is limiting \cite{Wooters82}, but in others it has shown new ways of tackling difficult computational tasks \cite{Ladd10}. In order to utilize quantum states for QIP, one needs to be able to effectively manipulate them. This is usually a daunting task, as any nontrivial manipulation tends to be accompanied by decoherence which deteriorates the quantum information of the states. An effective method of lessening the impact of decoherence employs the teleportation-based-QIP paradigm \cite{Gottesman99,Bartlett03,OBrien09}, in which the on-line operation is carried out deterministically with a help of a specifically prepared ancillary {\em resource} state, simple operations, measurement, and feedforward. In this way, the task of performing a universal quantum operation is translated to the task of generating a specific quantum state. This is usually much less of an issue, especially since the state can be, in principle, prepared by probabilistic means and then stored until it is needed.

In continuous variables (CV) quantum information processing, the currently readily available Gaussian operations \cite{OBrien09,Weedbrook12} allow us to straightforwardly prepare any Gaussian state - a state which can be described solely by Gaussian functions. However, in order to move out of this subset, a non-Gaussian operation is required. While none is available, which is deterministic, there is a probabilistic one, which has become a staple of CV quantum optics experiments. Single-photon detection, which can be physically implemented with help of single-photon on-off detectors, conditionally induces the states with a strongly non-Gaussian behavior \cite{Lvovsky01,Ourjoumtsev06TwoPhoton,Neergaard07}. When a series of photon subtractions (or, alternatively, additions) is accompanied by suitable displacements, it can be used to generate arbitrary superpositions of Fock states up to the number of subtractions used \cite{Dakna99,Fiurasek05}. Any finite energy quantum state can be, with any desired accuracy, realized as a finite superposition of Fock states if all the required features are obtained. This approach has been suggested for deterministic implementation of highly nonlinear weak cubic quantum gate \cite{Marek11}, which is a basic element of CV quantum information processing.

In the past, photon subtractions accompanied by displacements have been used to generate superpositions of zero and one photon \cite{Lvovsky02} and superpositions of up-to two photons \cite{Bimbard10}. These experiments, however, were carried out with pulsed lasers and are therefore not compatible with the current teleportation-based quantum operations \cite{Filip05,Yoshikawa07,Yoshikawa08,Miwa12}. This is because it is difficult to apply measurement and feedforward to those generated states which have a bandwidth of more than a few GHz. In this paper, we develop an experimental scheme using a continuous-wave (cw) laser as a light source to generate arbitrary superposition states with a bandwidth of 10 MHz. In particular, we generate superposition states of up-to three photons. The generated states are applicable to the current teleportation-based quantum operations \cite{Filip05,Yoshikawa07,Yoshikawa08,Miwa12}, and thus can be readily used to implement non-Gaussian gates. The generated states remarkably exhibit multiple negative areas of Wigner function, which can be not only exploited as better resource for CV quantum information processing \cite{Ralph03}, but also to completely characterize fundamental decoherence process of nonclassical states \cite{Deleglise08} from point of view of evolving system and environment.

The way the single-photon detections generate arbitrary superpositions of Fock states can be easily understood by considering an initial two-mode squeezed vacuum, which can be experimentally prepared by non-degenerate parametric process. It is expressed in the basis of photon number states $|n\rangle$ as
\begin{equation}
|\psi\rangle_{s,i} \propto \sum_{n}q^{n}|n\rangle_{s}|n\rangle_{i},
\end{equation}
where the characters $s$ and $i$ denote the signal and idler modes, respectively. The quantity $q$ $(0\le q < 1)$ depends on the pump power and the nonlinear coefficient of the nonlinear crystal. Linear optics is now used to split the idler mode into three, and to displace each of these modes $i_1$, $i_2$, and $i_3$ by coherent amplitudes $\beta_1$, $\beta_2$, and $\beta_3$. The idler modes are then measured by single-photon detectors and when the three-fold coincidence occurs, the signal mode is projected into the desired superposition state. In the limit of small pump power and small displacements, we can represent the projection process by
\begin{equation}\label{}
    |\psi\rangle_s \propto \langle 0|_i \left(\frac{a}{\sqrt{3}}+\beta_1\right)\left(\frac{a}{\sqrt{3}}+\beta_2\right)\left(\frac{a}{\sqrt{3}}+\beta_3\right)|\psi\rangle_{s,i},
\end{equation}
where $a$ is an annihilation operator acting on the idler mode, which represents the single-photon detection. The factor $\frac{1}{\sqrt{3}}$ arises from the splitting of the initial idler mode into the three separately measured modes. The output state then looks as
\begin{eqnarray}
\begin{split}
|\psi\rangle_s \propto\beta_{1}\beta_{2}\beta_{3}&|0\rangle_s +\frac{q}{\sqrt{3}}(\beta_{1}\beta_{2}+\beta_{2}\beta_{3}+\beta_{3}\beta_{1})|1\rangle_s \\
&+\frac{\sqrt{2}}{3}q^{2}(\beta_{1}+\beta_{2}+\beta_{3})|2\rangle_s +\frac{\sqrt{2}}{3}q^{3}|3\rangle_s .
\end{split}
\label{SuperpositionFomula}
\end{eqnarray}
We can see that generating arbitrary superpositions of Fock states from zero to three is only a matter of finding suitable values of the three displacement amplitudes. Similarly, with higher number of single-photon detectors, a superposition of higher Fock states would be viable.

\begin{figure}[htbp]
  \begin{center}
    \includegraphics[width=0.75\linewidth]{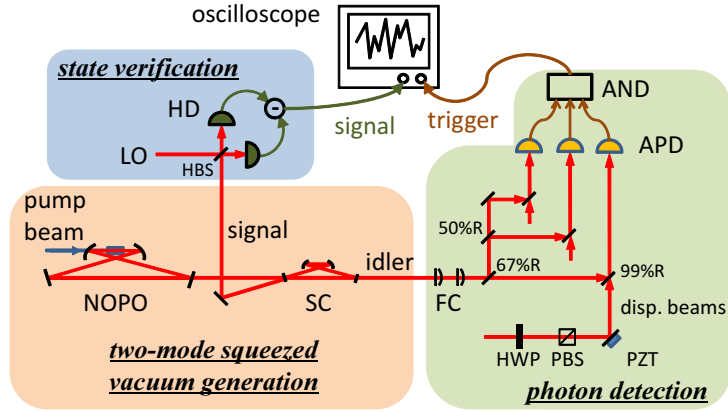}
  \end{center}
  \caption{Schematic of the experiment. The output of HD is recorded by a digital oscilloscope when a trigger is obtained. Triggers are obtained from an AND circuit when all of three APDs have clicks simultaneously. NOPO, non-degenerate optical parametric oscillator; SC, split cavity; FC, filter cavity; HD, homodyne detector; APD, avalanche photo diode; HBS, half beamsplitter; HWP, half-wave plate; PBS, polarization beamsplitter; PZT, piezo electric transducer.}
  \label{fig:setup.eps}
\end{figure}

\section{Experimental setup}
A schematic of the experiment is shown in Fig.~\ref{fig:setup.eps}. The light source is a cw Ti:Sapphire laser of 860 nm. In order to generate a two-mode squeezed vacuum, around 20 mW of pump beam of 430 nm is injected into a non-degenerate optical parametric oscillator (NOPO), which contains a periodically-poled $\mathrm{KTiOPO_{4}}$ crystal as an optical nonlinear crystal. The pump beam is generated by second harmonic generation of the fundamental beam, and frequency-shifted with an acousto-optic modulator by around 600 MHz (equal to free spectral range of NOPO, $\Delta\omega$). As a result, photon pairs of frequency $\omega$ (signal) and $\omega+\Delta\omega$ (idler) are obtained. The output photons are spatially separated by a split cavity whose free spectral range is $2\Delta\omega$. The photons of frequency $\omega+\Delta\omega$ passing through the split cavity are sent to two frequency filtering cavities \cite{Wakui07}, and are split into three beams with beamsplitters. Each beam is interfered with displacement beams at mirrors of 99\% reflectivity. Phase of displacement is controlled by piezo electric transducers, and amplitude of displacement is controlled by rotating half-wave plates followed by polarization beamsplitters. The idler beams are coupled to optical fibers to be sent to avalanche photo diodes (APDs, Perkin-Elmer, SPCM-AQRH-14 and SPCM-AQRH-16). The APDs output electronic pulses when they detect photons. The outputs are combined into an AND circuit to get three-fold coincidence clicks.

The signal beam is measured by homodyne detection with local oscillator beam of 10 mW. The homodyne current is sent to an oscilloscope and stored on every coincidence click. The density matrix and Wigner function of the output state are numerically reconstructed from a set of measured quadratures and phases of the local oscillator beam \cite{Smithey93,Lvovsky04}.

The presented experimental setup is capable of generating arbitrary superpositions of up-to three photons, simply by choosing a proper array of displacement parameters. To showcase this ability, we have generated a trio of quantum states, each of them strongly dependent on its three photon component.

\section{Results and discussions}
\begin{figure}[htbp]
  \begin{center}
    \includegraphics[width=\linewidth]{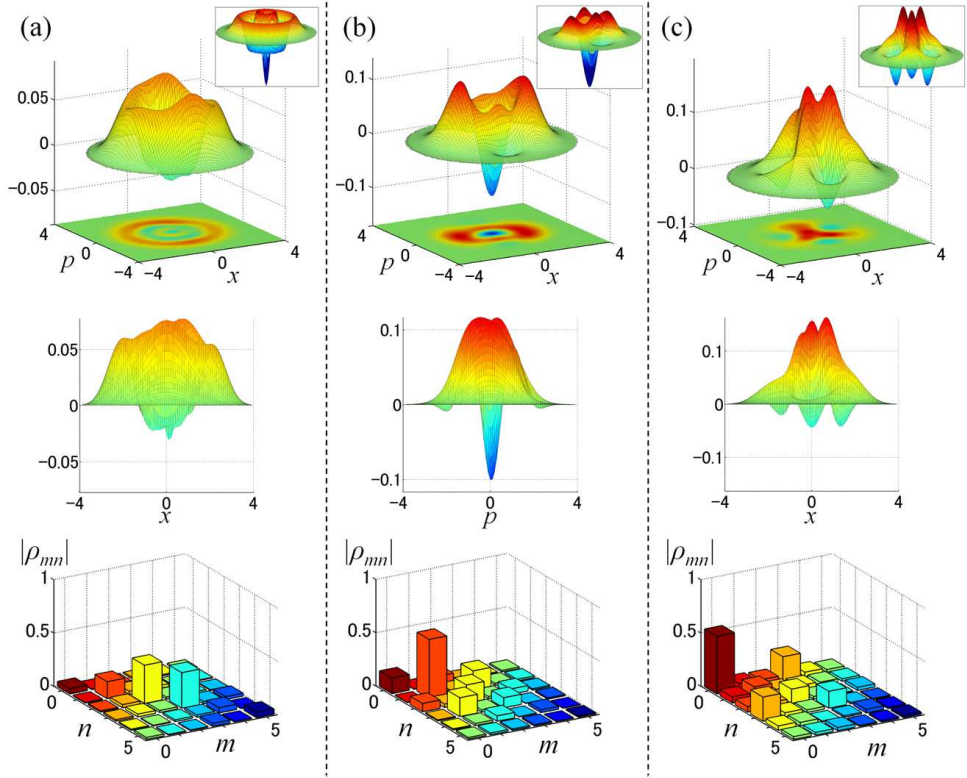}
  \end{center}
  \caption{The experimentally reconstructed density matrices and Wigner functions $(\hbar=1)$. (a) Three photon state, (b) coherent state superposition, (c) superposition of zero- and three-photon state with $s = 0.86 q$. The small insets of Wigner functions of the ideal states are shown for comparison.}
  \label{fig:AllResult.eps}
\end{figure}
At first, we generate the key resource, the three photon Fock state $|3\rangle$, which can be obtained without using any displacements. The defining features of this state are the distinctive presence of the three photon component, and the distinctive lack of presence of higher photon numbers. The density matrix reconstructed from 5000 data points is shown in Fig.~\ref{fig:AllResult.eps}(a) and it exhibits both the features mentioned. The three photon element $\rho_{33} = 0.33$ plays a significant role and the whole state is fairly well contained in the three photon subspace, with higher photon numbers populated only in 10 percent of the cases. Note that $\rho_{33}$ is equal to the fidelity of the state, which is defined as the overlap $F = \langle\psi|\rho_{\mathrm{exp}}|\psi\rangle$ of the experimentally generated state $\rho_{\mathrm{exp}}$ with the ideal state $|\psi\rangle$. The two photon and one photon contributions are caused by the experimental imperfections, such as optical losses and dark counts of the photon detectors, while the presence of higher photon numbers is caused by the strong pump power, which needed to be large enough to allow for a sufficient count rate (20 counts per minute).
Despite the imperfections, the Wigner function of the three photon state, also shown in Fig.~\ref{fig:AllResult.eps}(a), displays all the features one would expect from the three photon Fock state: it is spherically symmetrical and along any cut in the phase space it exhibits three distinctive regions of negativity.

The second generated state is the superposition of Fock states $|1\rangle$ and $|3\rangle$, which is achieved by using displacements $\beta_1 = -\beta_2 = \sqrt{2}q$ and $\beta_3 = 0$. For suitably selected parameters, this state is a good approximation of the coherent state superposition $|\mathrm{CSS}\rangle \propto |\alpha\rangle - |-\alpha\rangle$, which can play a very important role in quantum information processing \cite{Ralph03}. To be of use, the coherent state superposition needs to have a large enough amplitude. This comes with a very distinct feature: the wave function of a coherent state superposition has an infinite number of intersections with the horizontal axis. This has a consequence for the state's Wigner function - as the amplitudes of the coherent states grow, the number of regions of negativity should increase as well. In the past, approximations of odd superpositions of coherent states for travelling field of light were generated by squeezing a single photon (or by subtracting a photon from a squeezed state) \cite{Miwa12,Wakui07,Ourjoumtsev06,Neergaard06}. However, the states generated by this approach always have only a single region of negativity, no matter their apparent amplitude, which severely limits their potential applications. In order to obtain more regions of negativity and, consequently, more faithful approximations of coherent states superposition with higher amplitudes, one should employ higher photon numbers during the generation.

By doing exactly that, we have generated a state which is a good superposition of coherent state superposition with $\alpha = 1.3$. The fidelity with the ideal coherent state superposition was found to be $F =0.6$. However, the state is actually a squeezed coherent state superposition - subsequent antisqueezing could increase the amplitude to $\alpha = 1.6$ while simultaneously increasing the state fidelity to $F = 0.61$. In this sense, what was actually generated was the non-Gaussian keystone for a larger coherent state superposition \cite{Ourjoumtsev07, Menzies09}. Its density matrix and Wigner function, reconstructed from 10000 data points, can be seen in Fig.~\ref{fig:AllResult.eps}(b). The count rate is around 100 counts per minute. The three photon nature of the generated state is manifest in presence of three regions of negativity, which is exactly the number one would expect from a coherent state superposition with amplitude $\alpha = 1.6$. The presence of elements corresponding to Fock states 0 and 2 is again caused by losses at various stages of the experiment. By obtaining the multiple areas of negativity resulting from higher interference effects of the coherent states, we have reached a quality of state preparation previously obtained only for field in a cavity \cite{Deleglise08,Hofheinz09}.

The third generated state is the superposition of Fock states $|0\rangle$ and $|3\rangle$, which needs three different displacements during the state preparation stage, $\beta_{1}=s\mathrm{e}^{i\frac{\pi}{6}}$, $\beta_{2}=s\mathrm{e}^{i\frac{5\pi}{6}}$ and $\beta_{3}=s\mathrm{e}^{i\frac{3\pi}{2}}$, where $s$ is the displacement amplitude. Such the state is a good approximation of a different kind of coherent state superposition - $|\alpha\rangle + |\alpha e^{i \frac{2\pi}{3}}\rangle + |\alpha e^{-i\frac{2\pi}{3}}\rangle$, which can be seen as a sample qutrit state encoded in the coherent state basis. This coherent state basis is orthogonal in the limit of large $\alpha$, but similarly to the coherent state qubit basis, there is also a completely orthogonal basis formed of superpositions of Fock states invariant to $2\pi/3$ phase space rotation: $|0\rangle + \frac{\alpha^3}{\sqrt{6}}|3\rangle + \cdots$, $|1\rangle + \frac{\alpha^3}{2\sqrt{6}}|4\rangle +\cdots$, and $|2\rangle + \frac{\alpha}{2\sqrt{15}}|5\rangle+\cdots$. We have succeeded in generating the first of these basis states and the density matrix, reconstructed from 4000 data points, together with the Wigner function are shown in Fig.~\ref{fig:AllResult.eps}(c). The count rate is 50 counts per minute. We can see that the state is strongly nonclassical, with three areas of negativity, and that it possesses distinctive rotational $2\pi/3$ symmetry, which is exactly as predicted by the theory. The fidelity with the ideal state is $F=0.61$. Recently, an alternative procedure of similar state preparation for a field stored in a cavity has been suggested \cite{Raimond12}.

In all the presented results, there are minor contributions of photon number elements not agreeing with the idealistic expectations. Contributions of Fock states 4 and higher are generally caused by strong pumping, which was necessary in order to achieve a sufficient count rate. Undesirable photon number elements of less than three photons are caused by optical losses and dark counts of photon detectors. It should be pointed out, though, that all the states were reconstructed on a six dimensional Hilbert space without any loss correction. This is in stark contrast to previous work focused at generating superpositions of photon number states up to two \cite{Bimbard10}, where correction of 45\% loss was required to counteract low quantum efficiency of the detection. In our case, the high interference visibility of homodyne detection (97\%) and high quantum efficiency of photo diodes (99\%) add to the overall quantum efficiency for the whole experimental setup of 78\%. Consequently, the states could be reconstructed without loss correction and they are therefore suitable for use as ancillae in teleportation-based quantum operations in advanced CV quantum information processing.

\section{Conclusion}
We have constructed an experimental setup based on a cw laser, which is capable of generating superpositions of Fock states up to three. Since we used a cw laser as a light source, the generated states are compatible with teleportation-based CV quantum information processing \cite{Filip05,Yoshikawa07,Yoshikawa08,Miwa12}. We have tested the experimental setup by generating three characteristic states - the three photon Fock state to demonstrate the three-photon capability, and two superpositions to show that the nonclassical behavior can be also realized in a superposition. We have observed strong nonclassical features, manifesting in multiple areas of negativity, which were in good agreement with theoretical expectations, even without using any form of loss correction. This was made possible by high quantum efficiency of the experimental setup, which is indispensable for use in teleportation-based QIP. The scheme can also allow us to observe fundamental aspects of quantum decoherence of highly nonclassical states \cite{Deleglise08}, giving us access to both the evolving system and the environment. The experimental setup is not limited just to the preparation of the three states - arbitrary superpositions of Fock states of up-to three can be generated by using suitable array of displacements. This, in combination with high quantum efficiency of the setup and the cw platform ready for integration with an array of CV gates, makes this scheme a strong tool in the future CV non-Gaussian quantum information processing.

\section*{Acknowledgments}
This work was partly supported by PDIS, GIA, G-COE, APSA, and FIRST commissioned by the MEXT of Japan, and ASCR-JSPS. R.F. acknowledges support of P205/12/0577 of Czech Science Foundation. P.M. acknowledges support of P205/10/P319 of GACR.

\end{document}